\documentclass[a4paper,10pt]{article}
\usepackage{graphicx}
\usepackage{epsfig}
\usepackage[american]{babel}
\begin{document}

\title{Torus Fractalization and Intermittency}

\author{Sergey P.Kuznetsov}

\maketitle

\begin{center}
\emph{Institute of Radio-Electronics, Russian Academy of
Sciences,\\ Zelenaya 38, Saratov 410019, Russian
Federation}
\end{center}

\begin{abstract} The bifurcation transition is studied for the
onset of intermittency analogous to the Pomeau-Manneville
mechanism of type-I, but generalized for the presence of a
quasiperiodic external force. The analysis is concentrated on the
torus-fractalization (TF) critical point that occurs at some
critical amplitude of driving. (At smaller amplitudes the
bifurcation corresponds to a collision and subsequent
disappearance of two smooth invariant curves, and at
larger amplitudes it is a touch of attractor and repeller 
at some fractal set without coincidence.) For the TF
critical point, renormalization group (RG) analysis is developed.
For the golden mean rotation number a nontrivial fixed-point
solution of the RG equation is found in a class of
fractional-linear functions with coefficients depending on the
phase variable. Universal constants are computed responsible for
scaling in phase space ($\alpha=2.890053...$ and
$\beta=-1.618034...$) and in parameter space
($\delta_1=3.134272...$ and $\delta_2=1.618034...$). An analogy
with the Harper equation is outlined, which reveals important
peculiarities of the transition. For amplitudes of driving less
than the critical value the transition leads (in the presence of
an appropriate re-injection mechanism) to intermittent chaotic
regimes; in the supercritical case it gives rise to a strange
nonchaotic attractor. \end{abstract}


\section{Introduction}
It is commonly believed that under
parameter variation the turbulent dynamics in multi-dimensional
systems may arise via quasiperiodicity, in a course of subsequent
birth of oscillatory components with incommensurate frequencies,
followed by chaotization (see e.g., the works of Landau, Hopf, and
Ruelle and Takens [1,2,3]).

Now it is well known that actual details of the transition from
quasiperiodicity to chaos are very subtle and complicated. Some of
them can be revealed if we turn to a kind of restricted problem:
Suppose that the system may be decomposed to a master subsystem
with quasiperiodic behavior and a driven slave subsystem, and the
last can demonstrate transition to chaos. So, we may ask what are
possible scenarios of the onset of chaos in the second subsystem?
(Note an analogy with an approach to the problem of three bodies
in celestial mechanics: Being difficult in a general formulation,
it allows essential advance in a restricted version, under a
suggestion that one of the bodies is of negligible mass.) One of
the important results on this way of reasoning was a formulation
of the concept of strange nonchaotic attractor (SNA), which
typically appears in an intermediate region between order and
chaos [4-6]. In the phase space this is an object of fractal geometrical 
structure, but without instability in respect to the initial 
conditions in the driven system.

One more essential idea consists in application of the
renormalization group (RG) approach, proven to be very efficient
for understanding dynamics in critical states between order and
chaos (e.g. [7-16]). Starting with an evolution operator for some
definite time interval we are able to construct the evolution
operator for a larger interval. Then, we try to produce an
appropriate variable change to make the new operator as close as
possible to the original. This is just one step of the RG
transformation, and it may be repeated again and again to obtain
operators for larger and larger time intervals. As a result, we
arrive at some universal operator, which describes long-time
evolution of the system at the criticality. It is often
represented by a fixed point of the RG transformation. Studies of
this fixed-point operator together with consideration of its
relevant perturbations reveal properties of universality and
scaling for the transition. Originally, such an approach was
developed by Feigenbaum for the period-doubling scenario of the
onset of chaos [7,8]; afterwards it was applied to many other
situations, including quasiperiodicity at the chaos border [10-12]
and some cases of the birth of SNA [14-16].

Finally, we have to mention here a concept of intermittency suggested by
Pomeau and Manneville [17]. It occurs in very general circumstances near a
saddle-node bifurcation (also called 'tangent bifurcation', preferably in
the context of 1D maps). It has been studied in different aspects by many
authors. In particular, the RG approach has been applied to intermittency in
Refs. [18,19].

The goal of the present article is to consider a generalization of the
Pomeau-Manneville mechanism for the case of the presence of quasiperiodic
driving, and to reveal details of the bifurcation transition, which is an
analog of the tangent bifurcation in this case. It is natural to regard the
situation as one of possible scenarios of transition from quasiperiodicity
to chaos in the context of the mentioned "restricted problem". At small
amplitudes of driving, the transition is rather trivial and consists in
collision with coincidence (and subsequent disappearance) of a pair of
smooth stable and unstable tori, see e.g. [20, 21]. However, at a definite
value of the amplitude a non-trivial critical situation occurs. It allows
application of the RG approach that will be developed. Also, the associated
scaling properties will be revealed and discussed.

In Sec.II we introduce the basic model map and review its
dynamical phenomena in the presence of the external quasiperiodic
driving. 
In Sec.III we consider some details of dynamics in terms 
of rational approximations of the frequency parameter and 
locate numerically the critical point associated with threshold 
of fractalization at the moment of collision of the 
invariant curves.
In Sec.IV we discuss a
link between the problem under study and the Harper equation ---
the lattice version of the one-dimensional Schr\"odinger equation,
well known in the context of solid-state physics [22-27]. In Sec.V
the RG analysis is developed for the situation of tori
fractalization: The RG equation is derived, and results of its
numerical solution are presented. In Sec.VI we discuss scaling
properties of the dynamics at the critical point. In Sec.VII the
linearized RG equation is derived, the spectrum of eigenvalues is
obtained, and two relevant eigenvalues responsible for scaling in
the parameter plane are distinguished. In Sec.VIII we consider
consequences of these results concerning dynamics in a
neighborhood of the critical point in the parameter space. In
particular, we extract from the RG results the critical exponents
for the duration of the laminar stages of intermittency and
compare them with empirical numerical data. In conclusion we
discuss some perspectives of further studies in the context of the
general problem of understanding the transition from usual
quasiperiodic regimes ("smooth torus") to SNA and chaos.

\section{The model and basic phenomena}

Let us start with an example of quasiperiodically forced 1D map

\begin{equation}
\label{eq1} x_{n + 1} = f(x_{n} ) + b + \epsilon
\cos 2\pi nw,
\end{equation}

\noindent where $\epsilon$ and $w$ are the amplitude and frequency
parameters of the external force, respectively. We assume that the
frequency parameter, called also the rotation number, is taken to
be equal to the inverse golden mean, $w=(\sqrt {5} - 1)/2$. As to
the function $f(x)$, let us define it here as
\begin{equation}
\label{eq2} f(x) = \left\{\begin{array}{ll}
 x/(1-x), & x \le 0.75, \\
 9/2x-3, & x > 0.75. \\
\end{array} \right.
\end{equation}
(One branch of the mapping is
selected in a form of the fractional-linear function, $x/(1-x)$,
which appears naturally in analysis of dynamics near the tangent
bifurcation associated with intermittency, see e.g. [18,19]. The
other branch is attached somewhat arbitrarily to ensure presence
of the 're-injection mechanism' in the dynamics.)

At zero amplitude of driving what we have is a usual transition to chaos via
the Pomeau-Manneville intermittency of type I, controlled by parameter $b$, see
Fig.1(a). At $b < 0$ the map has two fixed points on the left branch, one
stable and one unstable. Under increase of $b$ these points approach one to
another, collide, and then disappear. After that, at $b > 0$, the narrow
"channel" remains at the place of former existence of the pair of the fixed
points, and travel across this channel is a slow process --  the laminar
stage of intermittency. Closer to the bifurcation point $b = 0$, larger the
number of iterations required to pass the channel. After visiting the
right-hand branch (the turbulent stage of the intermittency) the orbit
quickly returns to the left, and travels through the channel again and
again.

If the amplitude of driving is finite (although sufficiently
small), then, instead of the fixed points we observe a pair of
closed smooth invariant curves, attractor and repeller, see
Fig.1(b). (A closed invariant curve may be thought as a
cross-section of a torus. For brevity, it is convenient sometimes
to speak about stable and unstable tori rather than about the
invariant curves.) With increase of $b$ attractor and repeller
come nearer to each other and collide, and the localized
attractor-repeller pair disappears. After that an extended
attractor arises of a form shown in the right panel of Fig.1(b).
On the diagram a degree of darkness reflects relative duration of
presence of the orbit in different parts of the attractor.

While we remain close to the point of bifurcation, the laminar
stages of dynamics may be distinguished, which occupy an
overwhelming part of observation time, like in the case of the
usual Pomeau-Manneville intermittency. In Fig.1(b) they correspond
to a domain of the most long-living residence -- along the left
branch of the map, at the place of location of the former
attractor-repeller pair. In our study we will concentrate on the
analysis of the laminar stages in the same way as it is commonly
accepted in the case of conventional intermittency. For this, it
is sufficient to use the map 
\begin{equation} 
\label{eq3} 
x_{n +1} = x_{n}/(1 - x_{n}) + b + \epsilon \cos (2\pi nw).
\end{equation}

As the numerical simulations clearly demonstrate, the collision of
the attractor-repeller pair is of different nature at small and at
large amplitudes of driving. Similar observations were reported
earlier in computations for the driven circle map [21,28].

At $\epsilon$ less than some critical value $\epsilon _c$ (in our
map $\epsilon _c=2$) we observe that the
invariant curves remain smooth until the collision, and they
precisely coincide with one another at that moment (Fig.2(a)). For
$\epsilon = \epsilon _c$ 
they also coincide at the collision, but here the form 
of the invariant curve appears to be wrinkled (Fig.2(b)).
Finally, at $\epsilon > \epsilon _c$, 
one can see that the collision takes place only at some 
fractal subset of points on the invariant curves, and no 
coincidence of the entire curves is observed (Fig.2(c)).

The essential change in the nature of the transition with passage
from $\epsilon < \epsilon _c$ to $\epsilon > \epsilon_c$ may be
demonstrated also by computations of the Lyapunov exponent. Figure
3(a) shows the Lyapunov exponent as it behaves along the
bifurcation border (at the collision of the attractor-repeller
pair) as a function of the amplitude of driving. Observe that for
$\epsilon < \epsilon_c$ the Lyapunov exponent has constant zero
value, but for $\epsilon > \epsilon _c$ it becomes negative and
decreases with growth of $\epsilon$ according to a visually
perfect linear law. Figure 3(b) depicts a diagram for the Lyapunov
exponent dependence on $b$ at fixed $\epsilon = \epsilon _c$; it
shows essentially distinct behavior, apparently, a power law with
a non-trivial exponent.

The intriguing fact that the change of character of the
bifurcation in the model map (\ref{eq3}) takes place precisely at
$\epsilon _c = 2$ will be explained in Section IV.

\section{Method of rational approximations}

As is well known, the irrational $w = (\sqrt {5} - 1)/2$ taken as
the frequency parameter in the driven map is a limit of a sequence
of rationals $w_{k} = F_{k - 1}/F_{k}$, where $F_{k} $ are the
Fibonacci numbers ($F_{0} = 0,\,F_{1} = 1,\,F_{k + 1} = F_{k} +
F_{k - 1}$). Let us change the rotational number $w$ to its
approximant $w_{k}$, and introduce a parameter of initial phase
$u$ into the equation:
\begin{equation}
\label{eq4}
x_{n + 1} = x_{n}/(1 - x_{n}) + b + \epsilon \cos 2\pi
(nw_{k} + u).
\end{equation}

Now, below the transition, at any fixed $u$ we have a pair of
cycles of period $F_{k}$, one stable, and another unstable. The
Floquet eigenvalue, or multiplier $\mu$, which characterizes
decrease of a perturbation over one period of the stable cycle,
will depend on $u$. This dependence appears to be periodic, with
period $2\pi/F_k$. For a given $\epsilon$ we may select
numerically such $b$ that the maximal value of $\mu$ at some phase
reaches 1, and $\mu$ is less than 1 at other phases. It
corresponds just to the first tangent bifurcation, that is a
collision of the earlier stable cycle of period $F_{k}$ with its
unstable partner. Technically, the computations are simplified
with two observations: first, the maximum of the multiplier occurs
at $u = 0$, and, second, the initial condition for the cycle at
the situation of collision may be expressed explicitly (see Eq.
(\ref{eq15}) in Sec.IV).

In Table I we present numerical data for the values of $b$
corresponding to the cycle collision at $u = 0$ for the critical
amplitude $\epsilon = 2$. Figure 4 shows the multiplier as a
function of the phase $u$ in an interval of periodicity in a
moment of the first tangent bifurcation. The diagrams are plotted
at three subsequent levels of the rational approximation for the
amplitude parameter less, equal, and larger than 2.

For $\epsilon < 2$ the dependencies become more flat under
increase of the order of the rational approximation (the
bifurcation tends to become "phase independent"). In contrast, for
$\epsilon > 2$ the curves tend to become sharper. At $\epsilon =
2$ the form of the curves looks like stabilized at subsequent
levels of the rational approximation.

It is interesting to discuss a relation of these observations with
the behavior of the Lyapunov exponent at the transition. As we
consider attractor for the irrational rotation number in terms of
a certain rational approxinmant $w_{k} = F_{k - 1}/F_{k}$, it
looks like a collection (continuum set) of periodic orbits, each
of which is associated with a particular initial phase $u$ and has
a value of Lyapunov exponent $\Lambda (u) = \left( 1/F_{k}
\right)\ln \mu (u)$. To obtain in this approximation an estimate
for the Lyapunov exponent of the whole attractor, we have to
perform averaging over the initial phases, $\Lambda =
{\left\langle {\Lambda (u)} \right\rangle}$. From the behavior of
the multipliers in the subcritical case we conclude that the value
of $\Lambda$ will tend to zero under increase of the order of
rational approximation. In the supercritical case only a very pure
subset of the orbits will have multipliers distant from 0 and
close to 1, so, the average value of $\Lambda$ is negative. These
arguments are in agreement with the observed dependence of the
Lyapunov exponent on the parameter $b$ along the bifurcation curve
(see the previous section, Fig.3a).

As seen from Table I, the bifurcation sequence converges to a
well-defined limit, 
\begin{equation} 
\label{eq5} 
b = b_c = -0.597\,\,515\,\,185\,\,376\,\,121\,... 
\end{equation} 
(See also a remark in the final part of Sec.VI.)
It is a numerical estimate for the parameter value associated with the
critical point of fractalization of the colliding tori. It will be
referred to as the \textit{TF critical point}. Dynamics at this
point and in its vicinity is a main subject of our study in Sec.V
and further.

\section{A link with Harper equation}

A Schr\"odinger equation in a normalized form for a quantum
particle in a one-dimensional discrete lattice with additional
quasiperiodic potential reads 
\begin{equation}
\label{sch}
i{\frac{{\partial \psi _{n}} }{{\partial t}}} = \psi _{n + 1} +
\psi _{n - 1} - 2\psi _{n} + \left( {\epsilon \cos 2\pi nw}
\right)\psi _{n}, 
\end{equation} 
where $n$ is the spatial index,
$\epsilon$ is an amplitude of the quasiperiodic potential, and $w$
defines its wave-number. Alternatively, one can speak of a wave
process in a lattice medium with supplied sustained quasiperiodic
perturbation. For an oscillatory solution of frequency $\Omega$
(that corresponds to a state of quantum particle of definite
energy) the exponential substitution $\psi_{n} \propto \exp
\left( {i\Omega t} \right)$ yields 
\begin{equation} 
\label{eq9}
\psi _{n + 1} + \psi _{n - 1} - 2\psi _{n} + \left( {\Omega +
\epsilon \cos 2\pi nw} \right)\psi _{n} = 0.
\end{equation} 
This is the so-called Harper equation well-known in the context of
solid-state physics [22-27].

Let us return to our fractional-linear map (\ref{eq3}) and perform
a variable change 
\begin{equation} 
\label{eq6} 
x_{n} = 1 - \psi_{n}/\psi _{n - 1}.
\end{equation} 
The result is exactly the Harper
equation (\ref{eq9}) with $\Omega$ changed to $b$:
\begin{equation} 
\label{eq8} 
\psi _{n + 1} + \psi _{n - 1} - 2\psi
_{n} + \left( {b + \epsilon \cos 2\pi nw} \right)\psi _{n} = 0.
\end{equation}

The link between the Harper equation and the fractional-linear
mappings was noticed and exploited earlier by Ketoja and Satija
[25], although they were interested in some other problems than
that of our concern here. Recently the same idea was effectively
used for analysis of spectral properties of the Harper equation in
Ref.~[27].

At rational approximants of the wave-number $w$ the expression
(\ref{eq9}) becomes an equation with periodic coefficients.
Together with the Floquet condition 
\begin{equation} 
\label{eq10}
\psi _{n + q} = \mu \psi _{n} = \psi _{n} e^{i\tilde \beta q},
\quad \tilde \beta = (\arg \mu + 2\pi m)/q, \quad q = F_{k}
\end{equation} 
it gives rise to an eigenvalue problem: At any
given wave-number $\tilde{\beta}$ one can obtain (say, numerically
[24]) a spectrum of frequencies 
\begin{equation} 
\label{eq11}
\Omega = \Omega (\tilde \beta). 
\end{equation} 
It is called a dispersion equation for the waves in the medium governed by
Eq.(\ref{sch}). If the equation has a real root $\tilde \beta$ at
a given $\Omega$, it corresponds to wave propagation, or a
transmission band. If the equation has no real, but complex
solutions, we say that $\Omega$ is in a forbidden zone, or in a
non-transmission band. In this case no propagating waves, but
spatial exponential decay occurs at the given frequency.

To understand relation between nature of solutions of the Harper
equation and those of our original problem, let us turn to a
particular case of slow spatial variation of $\psi _{n}$, use the
continuous limit, and set $\epsilon = 0$. That yields
\begin{equation} 
\label{eq14} 
{\Psi} '' + b\Psi = 0{\rm .}
\end{equation} 
Now, at $b < 0$ we have solutions of the form $\Psi
_{n} = C\exp (\pm \sqrt {|b|} n)$, and, according to (\ref{eq14}),
$x_{n} = 1 - \psi _n /\psi _{n - 1} = 1 - \exp (\pm \sqrt {|b|})$.
It corresponds to presence of two fixed points. At $b > 0$ we
obtain $\Psi _{n} = C\cos (\sqrt {b} n)$. Then, $x_{n} = 1 - \psi
_{n}/\psi _{n - 1} \to \infty$ as $n \to \pi/(2\sqrt {b})$; it
means that no localized attractor is present. In the same manner,
a forbidden zone of the Harper equation must be associated with
existence of a localized attractor-repeller pair of the driven
fractional-linear map, while a transmission band corresponds to
presence of the "channel" and to the laminar stages of
intermittency (see also [27]).

It is easy to find that for $\epsilon = 0$ the transmission band
in the Harper model occupies an interval of $\Omega$ from 0 to 4.
At nonzero $\epsilon$ the forbidden zones ('gaps') arise inside
the band, and they become wider as $\epsilon $ grows. Under
increase of the order of rational approximation $k$, new and new
narrower gaps appear inside the transmission bands. It occurs that
for $\epsilon < 2$ in the limit $k \to \infty $ the transmission
bands of higher orders dominate over the forbidden zones (i.e.
they have a larger total width), but for $\epsilon > 2$ the
situation is opposite [23,24]. The transition from one type of
behavior to another is known as the localization-delocalization
transition. The structure of the transmission bands at the
transition appears to be a kind of Cantor-like set. Figure 5(a)
shows the transmission and forbidden zones colored, respectively,
by gray and white on the parameter plane $( \epsilon
,\,\,\Omega)$.

The fact that the transition in the Harper equation must take
place at $ \epsilon = 2 $ follows from the argument of Aubry
[29,24]. By means of the Fourier-like transformation
\begin{equation} 
\label{eq12} \phi _{k} = \hat {F}\psi _{n} =
{\sum\limits_{n = - \infty} ^{\infty} {\psi _{n} e^{2\pi inkw}}}
\end{equation} 
one obtains from (\ref{eq9}) an equation of similar
form 
\begin{equation} 
\label{eq12a} \phi _{k + 1} + \phi _{k - 1}
- 2\phi _{k} + ({\Omega} ' + {\epsilon }'\cos (2\pi ikw))\phi _{k}
= 0, 
\end{equation} 
but with parameters 
\begin{equation}
\label{eq13} {\Omega}'= 2 +
2(\Omega-2)/\epsilon,\,\,\,\,{\epsilon}' = 4/\epsilon.
\end{equation} 
Localization of a wave-function implies
delocalization of its transform, and vise versa. So, the
transition has to occur at $\epsilon = 2$ that corresponds to a
fixed point of the equation for $\epsilon$.

In Fig.5(a) one can find a forbidden zone in the bottom part of
the diagram. Obviously, its top border must correspond to the
threshold of intermittency in the fractional-linear map. As may be
observed from comparison of Fig.5(a) and (b), this is indeed the
case. The TF critical point found in the previous section
corresponds exactly to the lowest frequency associated with
appearance of the wave propagation at the localization --
delocalization transition in the Harper eguation (at $\epsilon =
2$).

The Harper equation (\ref{eq8}) possesses an evident symmetry
being invariant in respect to the spatial reflection $n \to - n$.
Hence, it is possible to construct a symmetric solution. For this
we have to set $\psi _{1} = \psi _{ - 1} = [1 - (b + \epsilon)/2]
\psi _0$. According to (\ref{eq6}), the respective orbit of
the fractional-linear map (\ref{eq3}) has an initial condition
\begin{equation} 
\label{eq15} 
x_0=(b + \epsilon)/(b +
\epsilon-2). 
\end{equation} 
If this orbit is localized (that
occurs at one special value of $b$ for each $\epsilon$), it will
correspond to the situation of the attractor-repeller collision.
As well, this is true for periodic orbits corresponding to
rational approximants $w_{k}$ at $u = 0$ (see Eq.(\ref{eq4})), and
this notion is technically useful for computation of the sequence
of parameter $b$ values converging to the critical point TF
(Sec.III).

\section{Renormalization group analysis}

Let us develop now the RG approach to the dynamics at the critical point TF. 
Here we prefer to write out the original model in a form of two-dimensional 
mapping

\begin{equation} 
\label{eq16} 
\begin{array}{l}
x_{n + 1} = x_{n}/(1 - x_{n} )+ b +
\epsilon \cos (2\pi u_n)),\\
u_{n+1}=u_n+w \pmod 1,
\end{array}
\end{equation}
and assume, for convenience, that the phase variable $u$
is defined in such way that it always belongs to the interval 
$(-0.5, 0.5)$.

As the frequency 
parameter is the inverse golden mean, it is natural to deal 
with the evolution operators corresponding to Fibonacci's 
numbers of iterations. 

We need to introduce here a new variable $X$ (it 
differs from $x$ by a $u$-dependent shift, 
but details will be explained below). 
Let $f^{F_{k}} (X,u)$ and $f^{F_{k + 1}} (X,u)$ 
are the functions representing transformation of $X$ after
$F_{k}$ and $F_{k + 1}$ iterations, respectively. 
To construct the next operator, for $F_{k + 2}$ steps, we start
from $(X,u)$ and perform first $F_{k + 1}$ iteration to arrive at
($f^{F_{k + 1} }(X,u),\,\,u + F_{k + 1} w$), and then the rest
$F_{k} $ iterations with the result
\begin{equation}
\label{eq17} f^{F_{k + 2}} (X,u) = f^{F_{k}} (f^{F_{k + 1}}
(X,u),\,\,u + wF_{k + 1} ).
\end{equation}

To have a reasonable limit behavior of the sequence of the
evolution operators we change scales for $X$ and $u$ by some
appropriate factors $\alpha$ and $\beta$ at each new step of the
construction, and define the renormalized functions as
\begin{equation}
\label{eq18} g_{k} (X,u) = \alpha ^{k}f(X/\alpha
^{k},\,\,(-w)^ku).
\end{equation}
Note that $wF_{k + 1} = -(-w)^{k + 1}\,\,(\,\bmod 1)$, so it is
natural to set $\beta=-1/w$. Rewriting (\ref{eq17}) in terms of
the renormalized functions we come to the functional equation
\begin{equation}
\label{eq19} g_{k + 2} (X,u) = \alpha ^{2}g_{k} (\alpha ^{- 1}g_{k
+ 1} (X/\alpha,\, -wu),\,\,w^{2}u + w).
\end{equation}

The same equation was obtained in the RG analysis of the critical
points TDT and TCT [15,16]. Here we will deal with some other
solution of that equation, associated with the FT critical point.
To find out of what kind this solution is, we may attempt to
compute the functions $g_{k} $ from direct iterations of the map
(\ref{eq16}).

As mentioned, the variable $x$ in the original map must be
distinguished from $X$ used in the derivation of the RG equation. In
other words, we have to produce a variable change to pass to an
appropriate 'scaling coordinate system' in the two-dimensional
phase space $(u,x)$. The new coordinates may be defined as
\begin{equation} 
\label{eq20} 
X \propto x - x_c + Pu +
Qu^2,\,\,\,U = u, 
\end{equation} 
where $x_c$ is obtained from
Eq.(\ref{eq15}) with substitution $\epsilon =\epsilon_c = 2,\,\,b
= b_c$, $P$ and $Q$ are some coefficients. 

To evaluate the coefficients $P$ and $Q$ we can act as follows. 
Let us perform iterations of the map (\ref{eq16}) 
at the critical point $\epsilon=2, b=b_c$, starting from $u_0=0$ 
and $x_0=x_c$ (see (\ref{eq15})), 
and compute the values 
of $u$ and $x$ after $F_k$ and $F_{k+1}$ iterations. Let they be
$(u_{F_k}, x_{F_k})$ and 
$(u_{F_{k+1}},x_{F_{k+1}})$, respectively. Three points $(0,x_c)$, 
$(u_{F_k}, x_{F_k})$, and $(u_{F_{k+1}},x_{F_{k+1}})$ determine 
a parabola on the $(u,x)$-plane, and its equation is given by 
$x-x_c+Pu+Qu^2=0$. 
The coefficients may be easily evaluated from coordinates 
of the three points. Of course, the result will depend on the level 
number $k$, and we must estimate the asymptotical limits for 
the coefficients; they are  $P = 5.92667$ and $Q = - 210.629$. 
(In fact, the convergence is rather slow, but it is possible to 
guess its character, and obtain sufficiently good estimates.)

Now the procedure
consists in the following:

\begin{itemize}
\item Fix $k$ and the respective $F_{k}$.

\item For given $X$ and $U$ define the initial conditions for the map
(\ref{eq16}): $x = XA\alpha ^{ - k} + x_c - PU - QU^{2},\,\,\,u =
U$ where $A$ is an arbitrary constant, $\beta=-1/w$, and 
$\alpha = 2.89$ (this value has been selected in a course of the computations 
as the most appropriate one).

\item Produce $F_{k} $ iterations of the map (\ref{eq16}).

\item Return to variables $(X,U)$ by the inverse change

$X = \alpha ^{k}A^{ - 1}\left( {x_{F_k} - x_c + Pu_{F_k} + Qu^2_{F_k}}
\right),\,\,U = u_{F_k}$. \end{itemize}

Figure 6 presents graphically a sample of results of such
computations for two Fibonacci numbers, $F_{k}=233$ and 377. The
3D plots of two obtained functions are superimposed; observe their
excellent agreement. (Yet better degree of coincidence was found
for larger Fibonacci numbers.) This is an indication that we deal
with a fixed-point solution of the functional equation
\begin{equation}
\label{eq21} g(X,u) = \alpha ^{2}g(\alpha ^{ - 1}g(X/\alpha,\, -
wu),\,\,w^{2}u + w).
\end{equation}

Now it is worth emphasizing that the maps determining evolution
over the Fibonacci numbers of iterations are constructed by a
repetitive application of the fractional-linear mappings, and,
hence, must relate to the same fractional-linear class. It implies
that we may search for solution of the Eq.(\ref{eq19}) in a form.
\begin{equation}
\label{eq22} g_{k} (X,u) = \frac{a_{k} (u)X + b_{k}(u)}{c_{k} (u)X
+ d_{k} (u)},
\end{equation}
where the coefficients $a$, $b$, $c$, $d$ are some functions of
$u$. Without loss of generality we may require them to satisfy to
an additional condition ('unimodularity')
\begin{equation}
\label{eq23} a_{k}(u)d_{k}(u) - b_{k}(u)c_{k}(u) \equiv 1,
\end{equation}
and set, as convenient, $c_{k} (0) = -1$. Substituting
(\ref{eq22}) into (\ref{eq19}) we arrive at the RG equation
reformulated in terms of the coefficients
\begin{equation}
\label{eq24}
\begin{array}{l}
\left( {\begin{array}{cc}
 a_{k+2}(u) & b_{k+2}(u) \\
 c_{k+2}(u) & d_{k+2}(u) \\
\end{array}} \right) =  \\

\left( {\begin{array}{cc}
 a_k(w^{2}u + w) & \alpha^2 b_k(w^{2}u + w) \\
 c_k(w^{2}u + w)/\alpha^2 & d_k(w^{2}u + w) \\
\end{array}} \right)
\left( {\begin{array}{cc}
 a_{k + 1}(-wu) & \alpha b_{k + 1} (-wu)\\
 c_{k + 1}(-wu)/\alpha & d_{k + 1}( - wu)\\
\end{array}}\right).
\end{array}
\end{equation}

To find the fixed-point of this functional equation numerically we
approximate the functions $a(u),\,\,b(u),\,\,c(u),\,\,d(u)$ by
finite polynomial expansions. (Actually, the representation via
Chebyshev's polynomials on an interval $u \in ( - 1,\,\,1)$ has
been used.) Then, we organize the RG transformation as a computer
program, which calculates the set of the expansion coefficients
for the functions $a_{k + 2} (u),\,\,b_{k + 2} (u),\,\,c_{k + 2}
(u)$ from two previous sets, $a_{k + 1}
(u),\,\,b_{k + 1} (u),\,\,c_{k + 1} (u)$ and $a_{k} (u),\,\,b_{k}
(u),\,\,c_{k} (u)$. (Note that due to the unimodularity, only
three of the four functions are independent.) The fixed-point
conditions are 
\begin{equation} 
\label{eq24a} 
(a_{k + 2} ,\,\,b_{k
+ 2} ,\,\,c_{k + 2} ) = (a_{k + 1} ,\,\,b_{k + 1} ,\,\,c_{k + 1} )
\,\,\, \mbox{and} \,\,\, (a_{k + 1} ,\,\,b_{k + 1} ,\,\,c_{k + 1}
) = (a_{k} ,\,\,b_{k} ,\,\,c_{k} ). 
\end{equation} 
In terms of the
polynomial representation it is equivalent to some finite set of
algebraic equations in respect to the unknown coefficients of the
polynomials and the unknown constant $\alpha$. This problem was
solved by means of the multidimensional Newton method. As an
initial guess, a function obtained from iterations of the original
map (see Fig.6) was used. The resulting coefficients for the
functions $a(u),\,\,b(u),\,\,c(u),\,\,d(u)$ corresponding to the
fixed point are presented in Table II, and graphically in
Fig.7(a). Figure 7(b) shows 3D plot for the fixed-point function;
it may be compared with Fig.6. The constant $\alpha $ that is the
scaling factor for $X$ variable, is found to be 
\begin{equation}
\label{eq25} 
\alpha = 2.890\,\,053\,\,525... 
\end{equation} 
in good agreement with the previously mentioned empirical estimate
$\simeq 2.89$.

It is worth mentioning one more universal constant associated 
with the critical point. 

Evaluating a derivative of $g(X,u)=(a(u)X+b(u))/(c(u)X+d(u))$ 
in respect to $X$ at the origin yields 
$\gamma = {\left[ {\partial g(X,u)}/{\partial X} \right]}_{X = 0,\,u = 0}
=1/{d(0)^2}=22.518745...$.

As $g(X,u)$ represents 
the asymptotic form of the evolution operator for Fibonacci's 
numbers of iterations, and $X$ differs from $x$ only by the $u$-dependent 
shift, we conclude that the constant $\gamma$ will appear as asymptotic 
value of the derivative ${\partial x_{F_k}}/{\partial x_0}$  
if $x_0$ is selected in accordance with (\ref{eq16}). 

It gives a foundation for a method of locating the critical point.
One composes a program to iterate the original map together with 
the recursive computation of ${\partial x_n}/{\partial x_0}$, and tries 
to select an appropriate value 
of $b$ to obtain ${\partial x_{F_k}}/{\partial x_0}=\gamma$. 
The result quickly converges to the critical point $b_c$
as $k$ grows. This method appears to be the most accurate, and the 
best numerical data (see (\ref{eq5})) have been obtained with its help.

\section{Scaling properties of dynamics at the critical point}

Let us consider attractor at the critical point of our model map
(\ref{eq16}). Its portrait is shown in Fig.8 in natural variables
$(u_{n},\,\,x_{n} )$. Depicting a part of the plot in
'scaling coordinates' $(U,\,X)$ we reveal a scaling property
intrinsic to the attractor: The structure is reproduced again and
again at each subsequent step of magnification by factors $\alpha$
and $\beta$ along the vertical and horizontal axes, respectively.
This scaling property follows directly from the fact that in
scaling coordinates the evolution operators for different
Fibonacci numbers of iterations are asymptotically the same, up to
the scale change (recall Fig.6).

From the scaling property one can deduce an asymptotic 
expression for the form of the invariant curve in small scales 
near the origin. The form reproduces itself under simultaneous 
scale change by factors $\alpha$ and $\beta=-1/w$ along the 
axes $X$ and $U$, respectively, so, it must behave locally as 
$X\propto|U|^\kappa$ with 
$\kappa={\log \alpha}/{\log |\beta|}\approx 2.2054$.
As follows, this is a smooth curve, twice differentiable at 
the origin, but the third derivative diverges. Due to ergodicity 
ensured by irrationality of the frequency, the weak singularity 
at the origin implies existence of the same type of singularities 
over the whole invariant curve, on a dense set of points. 
Apparently, the observed wrinkled form of the invariant curve at 
the critical point (see Figs.2 and 8) reflects a presence of 
the mentioned set of dense weak singularities.

Figure 9 shows evolution of Fourier spectra generated by the map
(\ref{eq16}) as we move in the parameter plane along the
bifurcation curve that corresponds to a threshold of
intermittency. These spectra may be useful for a comparison with
possible experimental studies of the transition. For small
$\epsilon$, i.e. far enough from the critical value, the spectrum
contains a few components. It is enriched by many additional lines
at intermediate frequencies as we come to the critical or
supercritical values of $\epsilon$. At the critical point the
spectrum has a self-similar structure. It can be revealed by the
use of the double logarithmic scale (as suggested in a different
context in Refs.[10-12]), see Fig.10.

\section{Linearized RG equation and spectrum of eigenvalues}

A shift of parameters in the map (\ref{eq16}) from the critical
point corresponds to some perturbation of the evolution operator,
and this perturbation will evolve under subsequent application of
the RG transformation (\ref{eq19}). Let us assume that the
perturbation retains our evolution operators in the class of
fractional-linear mappings. It means that we can search for
solution of Eq.(\ref{eq19}) in a form 
\begin{equation}
\label{eq26} 
g_{k}(X,u) = \frac{(a_{k} (u) + \tilde {a}_{k}
(u))X+b_{k} (u) + \tilde {b}_{k} (u)}{(c_{k} (u) + \tilde {c}_{k}
(u))X+d_{k} (u) + \tilde {d}_{k} (u)}, 
\end{equation} 
where $a(u),\,\,b(u),\,\,c(u),\,\,d(u)$ correspond to the fixed-point
solution, while the terms with tilde are responsible for the
perturbation. Then, a substitution $\left( {\tilde a_{k}
(u),\,\,\tilde b_{k} (u),\,\,\tilde c_{k} (u),\,\,\tilde d_{k}
(u)} \right) \propto \delta ^{k}$ gives rise to the eigenvalue
problem 
\begin{equation} 
\label{eq27} 
\begin{array}{ll}
 \delta ^{2}\left( {\begin{array}{cc}
 \tilde {a}(u) & \tilde {b}(u) \\
 \tilde {c}(u) & \tilde {d}(u) \\
\end{array}} \right) =
& \delta \left( {\begin{array}{cc}
 a(w^{2}u + w) & \alpha^2 b(w^{2}u + w) \\
 c(w^{2}u + w)/\alpha^2 & d(w^{2}u + w) \\
\end{array}} \right)
\left( {\begin{array}{cc}
 \tilde {a}(-wu) & \alpha \tilde {b}(-wu) \\
 \tilde {c}(-wu)/\alpha & \tilde {d}(-wu) \\
\end{array}} \right)\\
&+\left( {\begin{array}{cc}
 \tilde {a}(w^{2}u + w) & \alpha^2 \tilde {b}(w^{2}u + w) \\
 \tilde {c}(w^{2}u + w)/\alpha^2 & \tilde {d}(w^{2}u + w) \\
\end{array}} \right)
\left( {\begin{array}{cc}
 a(-wu) & \alpha b(-wu) \\
 c(-wu)/\alpha & d(-wu) \\
\end{array}} \right).\\
\end{array}
\end{equation}

As usual, only the eigenvalues larger than 1 in modulus may be of interest
because the respective perturbations grow under repetition of the RG
transformation and, hence, may influence the form of the long-time evolution
operators.

Numerically, we solved the problem by use of finite polynomial
approximations for the functions involved, with taking into
account the previously found fixed-point solution. The equation
(\ref{eq27}) then gives rise to an eigenvalue problem defined in
terms of finite-dimensional matrices acting in a vector space of
coefficients for the polynomial expansions of the functions
$\tilde {a}(u),\,\,\tilde {b}(u),\,\,\tilde {c}(u),\,\,\tilde
{d}(u)$, and it can be dealt with standard methods of linear
algebra. The computations reveal 13 eigenvalues larger or equal to
1 in modulus; they are listed in Table III.

Actually, only few of them are of relevance.

First, some of the found eigenvectors are associated with
infinitesimal variable changes. For example, with a substitution
$X \to X + \varepsilon $ in the map (\ref{eq22}) we we arrive at a
new map 
\begin{equation} 
X_{n+1}=\frac{a(u)X_{n} + \varepsilon
a(u) + b(u)}{c(u)X_{n} + \varepsilon c(u)+d(u)}-\varepsilon.
\end{equation} 
The right-hand part is represented in the first
order in $\varepsilon$ as 
\begin{equation} 
\label{eq28}
g(X,u)+\tilde g(X,u)\cong \frac{({a(u)+\tilde a(u)})
X_{n} + ({b(u)+\tilde b(u)})}{({c(u)+\tilde
c(u)}) X_{n} + ({d(u)+\tilde d(u)})},
\end{equation} 
where $(\tilde {a}(u),\,\,\tilde {b}(u),\,\,\tilde
{c}(u),\,\,\tilde {d}(u)) = \varepsilon \left( {1\, - c(u),\,\, -
d(u),\,\,1,\,\,0} \right)$. In a course of the RG transformation
$X$ is rescaled as $X \to X/\alpha$, and, respectively, the
renormalized relative coordinate shift is multiplied by $\alpha$
too. So, $\left( {1\, - c(u),\,\, - d(u),\,\,1,\,\,0} \right)$
represents an eigenvector, and $\alpha$ is the associated
eigenvalue. (One can verify it by a direct substitution of the
eigenvector into Eq.(\ref{eq27}); moreover, the assertion has been
checked also by accurate comparison of the present eigenvector
components with those found numerically as functions of $u$.) So,
we may exclude the eigenvalue $\alpha$ from the list, because a
perturbation of this kind in the evolution operator may be always
compensated by a shift of the origin. In the same way, we regard
several other eigenvectors as irrelevant, they are marked in Table
III as linked with variable changes.

Second, in the derivation of the RG equation we have used a definite order
for the efolution operators (first $F_{k + 1} $, then $F_{k} $ steps).
However, this order may be inverted, and it leads to a distinct alternative
formulation of the eigenvalue problem, namely,
\begin{equation}
\label{eq29}
\begin{array}{ll}
 \delta ^{2}\left( {\begin{array}{cc}
 \tilde {a}(u) & \tilde {b}(u) \\
 \tilde {c}(u) & \tilde {d}(u) \\
\end{array}} \right) =
& \delta \left( {\begin{array}{cc} \tilde a(-wu+w^{-1}) & \alpha
\tilde b(-wu+w^{-1}) \\
 \tilde c(-wu+w^{-1})/\alpha & \tilde d(-wu+w^{-1}) \\
\end{array}} \right)
\left( {\begin{array}{cc} a(w^2u) & \alpha^2 b(w^2u) \\
 c(w^2u)/\alpha^2 & d(w^2u) \\
\end{array}} \right)\\
&+\left( {\begin{array}{cc}
 a(-wu+w^{-1}) & \alpha b(-wu+w^{-1}) \\
 c(-wu+w^{-1})/\alpha & d(-wu+w^{-1}) \\
\end{array}} \right)
\left( {\begin{array}{cc} \tilde{a}(w^2u) & \alpha^2
\tilde{b}(w^2u)
\\
\tilde{c}(w^2u)/\alpha^2 & \tilde{d}(w^2u) \\
\end{array}} \right).\\
\end{array}
\end{equation}

The true evolution operators are in any case the multiple
compositions of the same original map. Thus, for the actual
perturbations the equations (\ref{eq27}) and (\ref{eq29}) must be
equivalent. In other words, only those eigenvectors may be of
relevance, which are common for both the eigenvalue problems. This
property was verified numerically for all found eigenvectors. As
observed, some of them do not satisfy the condition; in Table III
they are marked as relating to a non-commutative subspace.

The rest two eigenvalues
\begin{equation}
\label{eq30} 
\delta _{1} = 3.134\,\,272\,\,989\,...\,\,\,{\rm
and}\,\,\,\delta _{2} = w^{ - 1} = 1.618\,\,033\,\,979\,...
\end{equation}
are relevant and responsible for scaling properties of the
parameter space near the critical point TF.

If we depart from the critical point in the parameter plane along
the bifurcation curve of the attractor-repeller collision, the
first eigenvector appears not to contribute into the perturbation
of the evolution operator. In this case the only relevant
perturbation is associated with $\delta _{2}$. However, if we
choose a transversal direction, say, along the axis $b$, the
perturbation of the first kind appears. It means that a coordinate
system appropriate for observation of scaling in the parameter
plane has to be defined as shown in Fig.11. It is a curvilinear
system: one coordinate axis is the line $\epsilon = 2$, but
another follows the bifurcation border accounting its curvature.
In analytical expression it is sufficient to keep terms up to the
second order. (This is due to the concrete relation between
$\delta _1$ and $\delta _2$: $\delta _1 > \delta _2$ and $\delta
_1 > \delta_2^2 $, but $\delta _1 < \delta _2^3$; see other
examples of scaling coordinates for different critical points and
discussion of the role of the relation of the eigenvalues in Refs
[13,16,30,31].)

So, we set
\begin{equation}
\label{eq31}
\begin{array}{l}
 b = b_{c} + C_{1} + pC_{2} + qC_{2}^{2} , \\
 \epsilon = 2 + C_{2} , \\
 \end{array}
\end{equation}
where
\begin{equation}
\label{eq32} 
p= (2 - b_c)/4\cong -0.64938,\,\,q \cong - 0.33692.
\end{equation}

The expression for $p$ follows from the analogy with the Harper
equation and from the Aubry transformation rule (\ref{eq13}): An
infinitesimal shift of $\epsilon$ and $b$ along the tangent line
to the bifurcation border must correspond to the shift of
${\epsilon}'$ and ${b}'$ along the same line. The value of $q$ is
calculated numerically, from the curvature of the bifurcation
border.

Beside the obtained nontrivial solution of the RG equation there
exists also a trivial, phase independent fixed point
\begin{equation} 
\label{eq33} 
g(X,u) \equiv g(X) = X/(1-X),
\end{equation} 
with $\alpha = 1/w = 1.618034...$ Naturally, this
is the fixed point responsible for behavior on the subcritical
part of the bifurcation border and associated with the transition
accompanied by a collision of smooth invariant curves. The
eigenvalue problem for the linearized RG equation may be solved
analytically for this case, and it reveals a unique relevant
eigenvalue $\delta = 1/w^2=2.618034...$

\section{Dynamics in a neghborhood of the critical point \\ and
intermittency}

Let us discuss now a question on the peculiarities of
intermittency in the quasiperiodically forced map. First of all,
we outline a possibility of three distinct regimes at the onset of
intermittency:

\begin{itemize} \item subcritical, $\epsilon < \epsilon_c=2$,
when collision with coincidence of the smooth invariant curves 
(attractor and repeller) takes place at the moment of bifurcation; 
\item critical, $\epsilon=\epsilon_c$, it corresponds to
collision and coincidence of the wrinkled invariant curves 
(the threshold of fractalization);
\item supercritical,
$\epsilon > \epsilon_c$, there
collision of the invariant curves occurs at some fractal subset of points. 
\end{itemize}

Figure 12 (a)-(c) show the time dependencies for dynamical
variable generated by the model map (\ref{eq1}) just before and
after the transition for the mentioned three cases. In the
intermittent regimes the 'laminar stages' are interrupted by the
'turbulent bursts'. The laminar stages on the right panels
reproduce approximately the patterns of the left panels.

Relative duration of the laminar phases becomes larger as we
approach the transition point. In usual Pomeau - Manneville
intermittency of type I the average duration of the laminar stages
behaves as
\begin{equation}
\label{eq34} t_{\rm lam}\propto 1/|\Delta b|^\nu
\end{equation}
with $\nu = 0.5$ [17-19].

In presence of the quasiperiodic force the same law is valid in
the subcritical region $\epsilon < 2$. In the critical case
$\epsilon = 2$ the exponent is distinct. Indeed, as follows from
the RG results, to observe increase of a characteristic time scale
by factor $\theta = 1/w = 1.61803...$ we have to decrease a shift
of parameter $b$ from the bifurcation threshold by factor $\delta
_{1} = 3.13427\,...\,$. As follows, the exponent must be
\begin{equation} 
\label{eq35} 
\nu = \log \theta/\log \delta
_1\cong 0.42123.
\end{equation} 
(Note that substitution of $\delta$
factor associated with the trivial fixed point (\ref{eq33}) yields
just the result for the subcritical region, $\nu = 0.5$.)

Figure 13 shows the data of numerical experiments with the
fractional-linear map aimed to verify the theoretical predictions
for the exponent $\nu$. At each fixed $\epsilon$ we empirically
determined an average duration of passage through the 'channel' in
dependence on $\Delta b$ for ensemble of orbits with random
initial conditions and plotted the results in the double
logarithmic scale. For particular $\epsilon = 1.7$ (subcritical)
and 2 (critical) the dependencies fit the straight lines
of a definite slope. As seen from Table IV, the correspondence of
numerical results with the theoretical predictions is rather good.
At subcritical $\epsilon$ slightly less than 2 one can observe a
'crossover' phenomenon, that is the slope change from critical to
subcritical value at some intermediate value of $\Delta b$.

It is interesting that the results obtained for the supercritical region
also indicate a constant definite slope, $\nu \approx 0.45$. At this moment
it is not clear how to explain this observation theoretically.

Figure 14 shows diagrams for Lyapunov exponents versus parameter
$b$ for subcritical, critical, and supercritical constant values
of $\epsilon$ in the artificial map (\ref{eq1}), which includes
the reinjection mechanism.

For the subcritical case the intermittency threshold corresponds
to the onset of chaos: The Lyapunov exponent becomes positive
immediately after the transition.

In the supercritical region the Lyapunov exponent is yet negative
at the moment of the bifurcation, it cannot immediately become
positive, and the transition will be accompanied by a birth of an
SNA rather than a chaotic attractor.

In the critical situation the diagram demonstrates self-similarity
(at least, in the domain $b < b_c$): A magnification by factor
$\delta _{1} = 3.13427\,...\,$ along the horisontal axis, and by
factor $\theta = 1.61803...$ (the rescaling factor for time) along
the vertical axis gives rise to similar pictures.

Figure 15 demonstrates a chart of parameter plane, or the 'phase
diagram' in the natural variables (left panel) and in the scaling
coordinates (the right panel). The gray areas are those of
negative Lyapunov exponent, two distinct tones correspond to the
domains of existence of the localized attractors (smooth tori in
the bottom area) and to the intermittent regimes (the top area to
the right). Apparently, the last is the region of SNA. This
assertion may be deduced from arguments of Pikovsky and Feudel
[6]. Indeed, considering dynamics there in terms of rational
approximants one can notice that the phase-dependent bifurcations
will occur inevitably. In contrast, the white area of positive
Lyapunov exponent is the domain of chaotic intermittent regimes.
Figure 16 shows portraits of attractors at several representative
points of the phase diagram.

\section{Conclusion}

The present study was devoted to one special situation of transition from
conventional quasiperiodicity ('smooth torus') to chaos or SNA via
intermittency in a model map under quasiperiodic external driving with
frequency parameter defined as the inverse golden mean. 
The main attention was concentrated on a critical situation 
reached at one particular, sufficiently large amplitude of driving, 
associated with threshold of fractalization. Here the bifurcation 
transition analogous to the tangent bifurcation consists in 
collision with coincidence and subsequent disappearance of an 
attractor and a repeller represented by a pair of wrinkled 
invariant curves.
RG analysis appropriate for the
critical situation was developed, the fixed-point solution of the RG
equation was found in a class of fractional-linear functions, the constants
responsible for scaling in phase space and parameter space were computed.

Some related problems remain yet open, for example, concerning
global scaling properties and dimensions of the critical
attractor. Also a generalization for other irrational rotational
numbers is of interest. (The last seems undoubtedly possible
because of the analogy with Harper equation: there the criticality
at $\epsilon = 2$ occurs at arbitrary values of the rotational
number.) One more problem is a developement of an appropriate
approach to analysis of the transition in supercritical region,
which would be of the same significance, as the RG method is in the
critical and subcritical cases.

As is common in situations allowing the RG analysis, one can
expect that the quantitative regularities intrinsic to our model
map will be valid also in other systems relating to the same
universality class. In particular, it may be suggested that
transition to SNA observed in a quasiperiodically forced
subcritical circle map [28] is of the same nature. Also, it would
be significant to find this type of behavior in systems of higher
dimension, for example, in quasiperiodically driven invertible 2D
maps, which could represent Poincare maps of some flow systems.

It would be interesting to reveal details and regularities of
coexistence and subordination of the discussed type of critical
behavior with behaviors of distinct universality classes studied
in Refs. [14-16] (e.g. the torus-collision terminal and
torus-doubling terminal points).

\vspace{3mm} \noindent \textbf{{\Large Acknowledgements}}
\vspace{2mm}

\noindent The work has been supported by RFBR (grant No
00-02-17509). The appearance of this work was stimulated by
discussions with Arkady Pikovsky and Eireen Neumann during
workshop "Statistical Mechanics of Space-Time Chaos" organized by
Max Planck Institute of Physics of Complex Systems (Dresden, July
2000). I thank Igor Sataev for valuable help in numerical solution
of the RG equation and associated eigenvalue problem, and Andrew
Osbaldestin for discussions during my visit to Loughborough
University.

\newpage

\noindent \textbf{\Large{Figure captions}} \vspace{3mm}

\noindent \textbf{Figure 1.} Onset of chaos via the
Pomeau-Manneville intermittency of type I in the model map
(\ref{eq1}) at $\epsilon = 0$ (a), and the transition via
collision and disappearance of smooth closed invariant curves,
attractor and repeller (b). The diagrams are shown on the plane
$(x,{x}')$, where ${x}'$ relates to the moment of time one unit
later than $x$. The left panels correspond to situation before the
bifurcation, and the right panels to situation after the
transition.

\vspace{2mm} \noindent \textbf{Figure 2.} Bifurcation of collision
of the closed invariant curves in subcritical (a), critical (b),
and supercritical (c) situations, at $\epsilon = 1.7$, 2, and 2.3,
respectively. The invariant curves representing attractor are
shown by solid lines, and those for repeller by dashed. Parameter
$b$ grows from the left to the right, and the last panel in a row
corresponds to the moment of collision.

\vspace{2mm} \noindent \textbf{Figure 3.} The Lyapunov exponent at
the bifurcation border versus the amplitude parameter $\epsilon $
(a) and dependence of the Lyapunov exponent on parameter $b$ at
critical $\epsilon = 2$ (b) for the fractional-linear map with
quasiperiodic driving.

\vspace{2mm} \noindent \textbf{Figure 4.} Floquet eigenvalue, or
multiplier $\mu$ computed at three subsequent levels of rational
approximation versus phase variable $u$ at the values of $b$
corresponding to the moments of the first cycle collision for
subcritical (a), critical (b), and supercritical (c) amplitudes.

\vspace{2mm} \noindent \textbf{Figure 5.} Transmission bands
(gray) and forbidden zones (white) in the parameter plane $\left(
{\epsilon ,\,\Omega} \right)$ for the Harper equation (a), and the
chart of dynamical regimes (phase diagram) for intermittency under
quasiperiodic driving (b). The bifurcation curve of the
attractor-repeller collision on the panel (b) exactly corresponds
to the bottom border of the gray area in the diagram (a).

\vspace{2mm} \noindent \textbf{Figure 6.} 3D plots for the
functions obtained from direct iterations of the fractional-linear
map at $\epsilon = 2$, $b = b_{{\rm c}} = - 0.597515...$ with
appropriate renormalization as explained in text (the arbitrary
constant mentioned in the text is chosen as $A = 84$). The number
of iterations $F_{k} = 233$ and 377. A coincidence of both plots
indicate that the functions approach the fixed point of the RG
transformation.

\vspace{2mm} \noindent \textbf{Figure 7.} Coefficients for the
fixed-point fractional-linear solution of the RG equation versus
phase variable $u$ (a) and 3D plot of the universal function (b).

\vspace{2mm} \noindent \textbf{Figure 8.} Portrait of the critical
attractor in natural variables $u_{n} = nw + u,\,x_{n}$ (left
panel) and fragments of the picture shown under subsequent
magnification in the curvilinear 'scaling coordinate' system. At
each next step the magnification is increased by factor $\alpha =
2.89005...$ along the vertical axis and $\beta = - 1.61803...$
along the horizontal axis.

\vspace{2mm} \noindent \textbf{Figure 9.} Evolution of Fourier
spectra generated by the map (\ref{eq3}) along the bifurcation
curve that corresponds to a threshold of intermittency.

\vspace{2mm} \noindent \textbf{Figure 10.} Fourier spectrum at the
TF critical point presented in the double logarithmic scale.
Notice visible repetition of the structure in respect to a shift
along the frequency axis.

\vspace{2mm} \noindent \textbf{Figure 11.} Local coordinates on
the parameter plane of the fractional-linear map appropriate for
demonstrating of scaling.

\vspace{2mm} \noindent \textbf{Figure 12.} The dynamical variable
versus time in the model map (\ref{eq1}) just before and after the
transition: (a) a subcritical amplitude of driving, $\epsilon =
0.5$; (b) the critical case, $\epsilon = 2$; (c) a supercritical
case, $\epsilon = 2.3$.

\vspace{2mm} \noindent \textbf{Figure 13.} Data of numerical
experiments with the fractional-linear map: average duration of
passage through the 'channel' versus deflection from the
bifurcation threshold for several values of $\epsilon$ in the
double logarithmic scale. Observe a 'crossover' phenomenon, the
slope change from critical to subcritical value at some
intermediate value of $\Delta b$ for $\epsilon = 1.95$.

\vspace{2mm} \noindent \textbf{Figure 14.} Lyapunov exponent
versus parameter $b$ for subcritical, critical, and supercritical
constant values of $\epsilon$ in the map (\ref{eq1}). Illustration
of scaling for the critical case: insets are shown with subsequent
magnification by $\delta _{1} = 3.13427...$ along the horizontal
axis, and by factor $\theta = 1.61803...$ (the rescaling factor
for time) along the vertical axis.

\vspace{2mm} \noindent \textbf{Figure 15.} A chart of parameter
plane or the 'phase diagram' in natural variables (left panel) and
in scaling coordinates (the right panel). The gray areas
correspond to negative Lyapunov exponent values, with distinct
tones for localized attractors -- smooth tori in the bottom area,
and to intermittent regimes associated presumably with SNA in the
top area to the right.

\vspace{2mm} \noindent \textbf{Figure 16.} Portraits of attractors
at several representative points of the phase diagram.

\newpage

\begin{table}[!ht] \caption{The values of $b$ for the first
attractor-repeller collision at the critical amplitude
$\varepsilon = 2$} \vspace{3mm}
\begin{tabular}{|c|c|}
\hline 
$w_k$ & $b$ \\ \hline

8/13&-0.5989496730498198\\ \hline

13/21& -0.5993564164969890 \\ \hline

21/34& -0.5975700101088623 \\ \hline

34/55& -0.5977371349948819 \\ \hline

55/89& -0.5975077597293093 \\ \hline

89/144& -0.5975427315192966 \\ \hline

144/233& -0.5975125430279679 \\ \hline

233/377& -0.5975187094612125 \\ \hline

377/610& -0.5975146415227064 \\ \hline

610/987& -0.5975156490874847 \\ \hline

987/1597& -0.5975150899217102 \\ \hline

1597/2584& -0.5975152478940331 \\ \hline

2584/4181& -0.5975151698106684 \\ \hline

4181/6765& -0.5975151939749183 \\ \hline

6765/10946& -0.5975151829388339 \\ \hline

10946/17711& -0.5975151865779841 \\ \hline 
\end{tabular} 
\end{table}

\begin{table}[!ht] \caption{The polynomial coefficients for the
fixed-point solution of the RG equation} \vspace{3mm}

\begin{tabular}{|r|r|r|r|r|}
\hline 
&$a(u)$\hspace{7mm} & $b(u)$\hspace{7mm} & $c(u)$\hspace{7mm} &
$d(u)$\hspace{7mm} \\ \hline

$1$& 3.180070169& 0.329861441& -1.000000000& 0.210730746 \\ \hline

$u$ & -3.450688327& -0.003027254& -0.601533776& 0.167219763 \\
\hline

$u^{2}$ & -3.247090086& -8.330520346& 0.443802007& 3.062831633 \\
\hline

$u^{3}$ & 3.992032627& 9.444480715& 0.293604279& 1.865595582 \\
\hline

$u^{4}$ & 1.200194704& 11.133767028& -0.094027530& -1.597931152 \\
\hline

$u^{5}$ & -1.549680177& -13.205541847& -0.058193081& -1.044745269
\\ \hline

$u^{6}$ & -0.187890799& -4.313300571& 0.011160724& 0.358377518 \\
\hline

$u^{7}$ & 0.286983991& 5.592822816& 0.006427031& 0.223393531 \\
\hline

$u^{8}$ & 0.016720852& 0.734433497& -0.000847235& -0.044603872 \\
\hline

$u^{9}$ & -0.031115111& -1.104258820& -0.000457454& -0.025974096 \\
\hline

$u^{10}$ & -0.000950352& -0.069739548& 0.000044648& 0.003533563 \\
\hline

$u^{11}$ & 0.002215271& 0.125757870& 0.000022730& 0.001923598 \\
\hline

$u^{12}$ & 0.000035770& 0.004131194& -0.000001582& -0.000192702 \\
\hline

$u^{13}$ & -0.000110612& -0.009312156& -0.000000761& -0.000098574
\\ \hline

$u^{14}$ & -0.000000838& -0.000160051& & 0.000006910 \\ \hline

$u^{15}$ & 0.000003663& 0.000479736& & 0.000003335 \\ \hline

$u^{16}$ & & 0.000003822& & \\ \hline

$u^{17}$ & & -0.000016158& & \\
\hline 
\end{tabular} \end{table}

\begin{table}[!ht] \caption{The eigenvalues larger or equal to 1
in modulus} \vspace{3mm}

\begin{tabular}{|r|c|l|}
\hline 
Eigenvalue & Designation & Interpretation \\ \hline

$3.134272989$ & $\delta_1$ & Relevant eigenvalue \\ \hline

$-2.890053625$ & $-\alpha$ & Non-commutative subspace \\ \hline

$2.890053625$ & $\alpha$ & The variable change $x \leftarrow x +
\varepsilon$ \\ \hline

$-1.786151370$ & $-\alpha w^{-1}$ & The variable change $x
\leftarrow x + \varepsilon u$ \\ \hline

$1.786151370$ & $\alpha w^{-1}$ & Non-commutative subspace \\
\hline

$-1.618033979$ & $-w^{-1}$ & The variable change $u \leftarrow u +
\varepsilon$ \\ \hline

$1.618033979$ & $\delta_2=w^{-1}$ & Relevant eigenvalue \\ \hline

$1.618033979$ & $w^{-1}$ & Violation of the unimodularity \\
\hline

$1.103902257$& $\alpha w^{-2}$ & The variable change $x \leftarrow
x + \varepsilon u^2$ \\ \hline

$-1.103902257$ & $-\alpha w^{-2}$ & Non-commutative subspace \\
\hline

$1.000000000$ & $1$ & The variable change $x \leftarrow
x(1+\varepsilon )$ \\ \hline

$-1.000000000$ & $-1$ & Non-commutative subspace \\ \hline

$-1.000000000$ & $-1$ & Non-commutative subspace \\
\hline 
\end{tabular} \end{table}

\begin{table}[!ht] \caption{Comparison of the numerical results
and RG predictions for the critical exponent} \vspace{3mm}

\begin{tabular}{|c|c|c|}

\hline 
$\varepsilon$ & $\nu$, numerics & $\nu$, theory \\ \hline

$1.7$ & $0.508$ & $0.5$ \\ \hline

$2.0$ & $0.424$ & $0.42123$ \\ \hline

$2.2$ & $0.452$ & ? \\ \hline

$2.3$ & $0.456$ & ? \\

\hline 
\end{tabular} \end{table}

\end{document}